\documentstyle[12pt,aps]{revtex}
\def\be{\begin{equation}}
\def\ee{\end{equation}}
\def\ba{\begin{array}}
\def\ea{\end{array}}
\def\bea{\begin{eqnarray}}
\def\eea{\end{eqnarray}}
\def\nn{\nonumber}
\def\cD{\cal D}
\def\cL{\cal L}
\def\cB{\cal B}
\def\cK{\cal K}
\def\cLd{{\cal L}^{\dagger}}

\def\sta{\sin\theta}
\def\cta{\cos\theta}
\def\sda{\sin^2\theta}

\def\coa{\cot\theta}
\def\sqd{\sqrt{2}}

\def\p{\partial}
\def\pr{\p_r}
\def\pv{\partial_v}
\def\pta{\p_{\theta}}
\def\pvi{\p_{\varphi}}

\def\spr{\frac{\p}{\p r_*}}

\def\spdr{\frac{\p^2}{\p r_*^2}}
\def\spdvr{\frac{\p^2}{\p r_* \p v_*}}
\def\spdra{\frac{\p^2}{\p r_* \p \theta_*}}
\def\spdri{\frac{\p^2}{\p r_* \p \varphi_*}}

\begin{document}
\baselineskip 21.8pt

\title{Hawking Radiation of Dirac Particles in an
Arbitrarily Accelerating Kinnersley Black Hole}

\author{S. Q. Wu\thanks{E-mail: sqwu@iopp.ccnu.edu.cn}
and X. Cai\thanks{E-mail: xcai@ccnu.edu.cn}}
\address{Institute of Particle Physics, Hua-Zhong
Normal University, Wuhan 430079, P.R. China}
\maketitle
\date{}

\vskip 1.5cm
\begin{abstract}
Quantum thermal effect of Dirac particles in an arbitrarily accelerating
Kinnersley black hole is investigated by using the method of generalized
tortoise coordinate transformation. Both the location and the temperature
of the event horizon depend on the advanced time and the angles. The Hawking
thermal radiation spectrum of Dirac particles contains a new term which
represents the interaction between particles with spin and black holes
with acceleration. This spin-acceleration coupling effect is absent from
the thermal radiation spectrum of scalar particles.

\end{abstract}

\newpage
\section{Introduction}

It has been more than a quarter century since Hawking's remarkable
discovery \cite{Hawk} that a black hole is not completely black but
can emit radiation from its event horizon. An important subject in
black hole physics is to reveal the thermal properties of various
black holes. Last few decades has witnessed much progress on investigating
the thermal radiation of scalar fields or Dirac particles in some
stationary axisymmetry black holes. Nevertheless most of these efforts
(see \cite{DR,Xuetc,WC}, for examples) were concentrated on studying
the thermal properties of static or stationary black holes. Because
a realistic black hole in astrophysics can radiate or absorb matter
surrounding it, it is nonstationary and evolves in the time. Thus the
thermal properties of non-stationary spacetimes are more interesting
than that of static or stationary black holes, and worth much more
studies about them. A well-known method to determine the location and
the temperature of the event horizon of a dynamical black hole is to
calculate vacuum expectation value of the renormalized energy momentum
tensor \cite{HB}. But this method is very complicated, it gives only an
approximate value of the location and that of the temperature. Thus it
is of limited use and meets great difficulties in many cases.

To study the Hawking evaporation of the non-stationary black holes,
Zhao and Dai \cite{ZD} suggested a new method of the generalized
tortoise coordinate transformation (GTCT) which can give simultaneously
the exact values of the location and the temperature of the event
horizon of a non-stationary black hole. By generalizing the common
tortoise-type coordinate $r_* = r +\frac{1}{2\kappa}\ln(r -r_H)$ in
a static or stationary spacetime \cite{DR,San} (where $\kappa$ is
the surface gravity of the studied event horizon) to a similar form
in a non-static or non-stationary spacetime \cite{ZD} and allowing
the location of the event horizon $r_H$ to be a function of the advanced
time $v = t +r_*$ and/or the angles $\theta, \varphi$, the GTCT method
reduces Klein-Gordon or Dirac equation in a known black hole spacetime
to a standard wave equation near the event horizon. For instances,
the location of the event horizon is a constant ($r_H = 2M$) in the
Schwarzschild black hole while it is a function of the advanced time
($r_H = r_H(v)$) in a Vaidya-type space-time. This method has been
applied to investigate the thermal radiation of scalar particles in
the non-uniformly accelerating Kinnersley ``photon rocket'' solution
\cite{LZZZ} and in the non-uniformly accelerating Kerr black hole
\cite{WZSZZ} as well.

However, it is very difficult to investigate the quantum thermal effect
of Dirac particles in the non-stationary black holes. The difficulty lies
in the non-separability of variables for the Chandrasekhar-Dirac equation
\cite{CD} in the most general spacetimes. The Hawking radiation of Dirac
particles has been studied only in some non-static black holes \cite{LZWMY}.
Recently we \cite{WUCAI} have tackled with the evaporation of Dirac particles
in a non-stationary axisymmetric black hole. Making use of the GTCT method,
we consider the asymptotic behaviors of the first-order and second-order
forms of Dirac equation near the event horizon. Using the relations between
the first-order derivatives of Dirac spinorial components, we eliminate the
crossing-terms of the first-order derivatives in the second-order equations
and recast each equation to a standard wave equation near the event horizon.
Not only can we re-derive all results obtained by others \cite{KGNK}, but also
we find that the Fermionic spectrum of Dirac particles displays another new
effect dependent on the interaction between the spin of Dirac particles and
the angular momentum of black holes. This spin-rotation effect is absent from
the Bosonic spectrum of Klein-Gordon particles.

It is natural to see whether or not our method can work effective in other
cases. In this paper, we apply it to deal with the Hawking effect of Dirac
particles in a non-spherically symmetric and non-stationary Kinnersley black
hole, namely Kinnersley ``photon rocket'' solution \cite{Kin}. The local event
horizon of a dynamical black hole is determined here by the null hypersurface
condition. We use different methods to deduce the location of the event horizon
and find that they all give the same result. By means of a GTCT, we can also
derive the event horizon equation from the limiting form of the first-order
Dirac equation near the event horizon. The location and the shape of the
Kinnersley black hole is not spherically symmetric \cite{LZZZ}. Then we turn
to the second-order Dirac equation. With the aid of a GTCT, we adjust the
temperature parameter in order that each component of Dirac spinors satisfies
a simple wave equation after being taken limits approaching the event horizon.
We demonstrate that both the shape and the temperature of the event horizon
of Kinnersley black hole depend on not only the time, but also on the angles.
The location and the temperature coincide with those obtained by investigating
the Hawking effect of Klein-Gordon particles in the accelerating Kinnersley
black hole \cite{LZZZ}. But the thermal radiation spectrum of Dirac particles
shows a new effect dependent on the interaction between the spin of Dirac
particles and the angular acceleration of black holes. This effect displayed
in the Fermi-Dirac spectrum is absent in the Bose-Einstein distribution of
Klein-Gordon particles. We find that this spin-acceleration coupling effect
does not exist in a non-uniformly rectilinearly accelerating Kinnersley black
hole.

The paper is outlined as follows: In Sec. 2, we introduce the most
general GTCT and derive the equation that determines the location of
the event horizon from the null surface condition. Sec. 3 is devoted
to discussing the Hawking radiation of Dirac particles in the Kinnersley
spacetime. First, we work out the explicit form of Dirac equation in
the Newman-Penrose \cite{NP} formalism and investigate the asymptotic
behavior of the first-order Dirac equation near the event horizon. The
equation that determines the location of the event horizon can be
inferred from the vanishing determinant of the coefficients of the
first-order derivative terms. Next, we use the relations between the
first-order derivative terms to eliminate the crossing-term of the
first-order derivatives in the second-order Dirac equation near the
event horizon, and adjust the parameter $\kappa$ introduced in the
GTCT so as to recast each second-order equation into a standard wave
equation near the event horizon. In the meantime, we can get an exact
expression of the Hawking temperature. Then the second-order equation
is manipulated by separation of variables and the thermal radiation
spectrum of Dirac particles are obtained by Damour-Ruffini-Sannan's
method \cite{DR,San}. In Sec. 4, we give a brief discussion about the
new effect which represents the interaction between the spin of particles
with spin-$1/2$ and the angular acceleration of black holes.

\section{Kinnersley black hole and its event horizon}

The Kinnersley metric \cite{Kin}, generally called as the ``photon
rocket'' solution, is interpreted as the external gravitational field
of an arbitrary accelerating mass. In the advanced Eddington-Finkelstein
coordinate system $[v; r; \theta; \varphi]$, the line element of
Kinnersley's rocket solution reads
\be
ds^2 = 2dv\big(G dv -dr -fr^2 d\theta +gr^2\sin^2\theta
d\varphi\big) -r^2\big(d\theta^2 +\sin^2\theta d\varphi^2\big) \, ,
\label{rocket}
\ee
where $2G = 1 -2M(v)/r -2ar\cta -r^2W^*W$, $W = f-ig\sta$, $W^* = f
+ig\sta$, $f = b\sin\varphi +c\cos\varphi -a\sta$, and $g = (b\cos\varphi
-c\sin\varphi)\coa$. The metric (\ref{rocket}) is of type-D under
Petrov classification. The arbitrary function $M(v)$ describes the
change in the mass of the source as a function of the advanced time;
$a = a(v)$, $b = b(v)$ and $c = c(v)$ are acceleration parameters: $a$
is the magnitude of acceleration, $b$ and $c$ are the rates of change
of its direction. The co-moving spherical coordinate system is oriented
in such a way that the direction $\theta = 0$ to the north pole always
coincides with the direction of the acceleration.

The spacetime geometry of an evaporating black hole is characterized by
three surfaces: the timelike limit surface, the apparent horizon and the
event horizon. According to York \cite{York}, the horizons of a dynamical
black hole may be obtained to first order in luminosity by note that (i)
the apparent horizons are the outermost ``trapped'' surfaces that the
expansion of null-geodesic congruences (or null rays parametrized by $v$)
$\vartheta \approx 0$; (ii) the event horizons are null surfaces where
the acceleration of null-geodesic congruences $d^2r/dv^2 \approx 0$, or
equivalently they are determined via the Raychadhuri equation by the
requirement that $d\vartheta/dv \approx 0$ as they must be strictly null,
and (iii) the timelike limit surfaces are defined as surfaces such that
$g_{vv} = 0$.

It is generally accepted that the event horizon is necessarily a null surface
and is defined by the outermost locus traced by outgoing photons that can
``never'' reach arbitrarily large distances \cite{York}. In a nonstationary
black hole spacetime, the event horizon should still be a null surface that
satisfies the null surface condition: $g^{ij}\p_i F\p_j F = 0$, and the event
horizon determined by the above null hypersurface condition is, in fact, a
local event horizon. We shall adopt this definition and use different methods
to derive the equation that determines the location of local event horizon of
an arbitrarily accelerating Kinnersley black hole. We find that each method
can give the same result consistently.

First, let's seek the local event horizon of Kinnersley spacetime
(\ref{rocket}) by using of the null surface condition. From the null
surface equation $F(v,r,\theta,\varphi) = 0$, namely
$r = r(v,\theta,\varphi)$, one can easily obtain
\be
\p_v F +\p_r F\p_v r = 0 \, , ~~~~\pta F +\p_r F\pta r = 0 \, ,
~~~~\pvi F +\p_r F\pvi r = 0 \, . \label{pde}
\ee
Substituting (\ref{pde}) into the explicit expression of the null surface
condition $g^{ij}\p_i F\p_j F = 0$ in the Kinnersley metric (\ref{rocket})
\bea
&&\big(2G +r^2W^*W\big)(\pr F)^2 +2\pr F\big(\pv F
-f\pta F +g\pvi F\big) \nn \\
&&~~~~~~~~~~~~~~~~~~~~~~~~~
+\frac{1}{r^2}(\pta F)^2 +\frac{1}{r^2\sda}(\pvi F)^2 = 0 \, ,
\label{nc}
\eea
one gets
$$2G +r^2W^*W -2\p_v r +2f\pta r -2g\pvi r +\frac{(\pta r)^2}{r^2}
+\frac{(\pvi r)^2}{r^2\sda}  = 0 \, . $$
The local event horizon is the hypersurface $r = r_H(v,\theta,\varphi)$
that satisfies the above equation or
\bea
&&1 -\frac{2M}{r_H} -2ar_H\cta -2r_{H,v} +2fr_{H,\theta}
-2gr_{H,\varphi} \nn\\
&&~~~~~~~~~~~~~~~~~~~~~~~~~
+\Big(\frac{r_{H,\theta}}{r_H}\Big)^2
+\Big(\frac{r_{H,\varphi}}{r_H\sta}\Big)^2 = 0 \, , \label{loca}
\eea
in which $r_{H,v} = \p_v r_H$, $r_{H,\theta} = \pta r_H$ and $r_{H,\varphi}
= \pvi r_H$ can be viewed as parameters depicting the evolution of the event
horizon.

When $a = b = c = 0$ but $M\not= 0$, the event horizon of Vaidya black
hole is located at $r_H = 2M/(1 -2r_{H,v})$; When $M = b = c = 0$ and
$a = const$, the Rindler event horizon of the uniformly rectilinearly
accelerating observer satisfies
$$1 -2ar_H\cta -2ar_{H,\theta}\sta +\frac{r_{H,\theta}^2}{r_H^2} = 0 \, ,$$
it is a paraboloid of revolution $r_H = 1/a(\cta\pm 1)$.

In the case where $M(v)$, $a(v)$, $b(v)$ and $c(v)$ are not equal to zero,
Eq. (\ref{loca}) has in general three roots. In this general case, the
analysis is a little involved, and will not be discussed here. There should
exist two kinds of event horizon: Rindler-type horizon and Schwarzschild-type
horizon. All of them depend not only on $v$, but also on $\theta$, $\varphi$.
It means that the location of the event horizon and the shape of the black
hole change with time. The location of event horizon is in accord with that
obtained in the case of discussing about the thermal effect of Klein-Gordon
particles in the same spacetime \cite{LZZZ}.

Next, we adopt the GTCT method to deduce the equation of local event horizon.
As the Kinnersley metric is lack of any symmetry, we introduce the most general
form of the GTCT \cite{ZD} as follows
\bea
&&r_* = r +\frac{1}{2\kappa}\ln \big(r -r_H\big) \, ,
~~~~ v_* = v -v_0 \, , \nn\\
&&\theta_* = \theta -\theta_0 \, ,
~~~~~~~~~~~~~~~~~~~~~ \varphi_* = \varphi -\varphi_0 \, ,
\label{trans}
\eea
namely,
\bea
&&dr_* = dr +\frac{1}{2\kappa(r -r_H)}\Big(dr -r_{H,v} dv
-r_{H,\theta}d\theta -r_{H,\varphi}d\varphi\Big) \, , \nn\\
&&dv_* = dv \, , ~~~~~~~~~~~~~~~~ d\theta_* = d\theta \, ,
~~~~~~~~~~~~~~~~ d\phi_* = d\varphi \, , \nn
\eea
where $r_H = r_H(v,\theta,\varphi)$ is the location of the event horizon,
$\kappa$ is an adjustable parameter and is unchanged under tortoise
transformation. All parameters $v_0, \theta_0$ and $\varphi_0$ are
arbitrary constants characterizing the initial state of the hole.

Applying the GTCT (\ref{trans}) to the null hypersurface equation
(\ref{nc}) and taking the $r \rightarrow r_H(v_0,\theta_0,\varphi_0)$,
$v \rightarrow v_0$, $\theta \rightarrow \theta_0$ and $\varphi
\rightarrow \varphi_0$ limits, the event horizon equation is then
obtained by letting the term in the bracket before $(\spr F)^2$
to be zero, \footnote{Throughout the paper, we make a convention
that all coefficients in the front of each derivatives term take
values at the event horizon $r_H = r_H(v_0,\theta_0,\varphi_0)$
when a GTCT is made and followed by taking limits approaching the
event horizon, e.g., $G$, $f$, $g$ and $W$ take their values
at $v = v_0$, $\theta = \theta_0$ and $\varphi = \varphi_0$.}
\be
2G -2r_{H,v} +r_H^2W^*W  +2f r_{H,\theta} -2g r_{H,\varphi}
+\frac{r_{H,\theta}^2}{r_H^2}
+\frac{r_{H,\varphi}^2}{r_H^2\sda_0} = 0 \, .
\label{eh}
\ee
Eq. (\ref{eh}) is just the same equation (\ref{loca}) when $v = v_0$,
$\theta = \theta_0$ and $\varphi = \varphi_0$. Because here we deal with
the case of a slow evaporation of black holes, we need later only consider
the situation very close to the initial state of the event horizon, namely
$r_H \approx r_H(v_0,\theta_0,\varphi_0)$ when $v \approx v_0$, $\theta
\approx \theta_0$ and $\varphi \approx \varphi_0$. This assertion is
due to that the GTCT approach is a local analysis method, the latter
originating from that Hawking radiation comes from vacuum fluctuations
near the event horizon.

\section{Hawking evaporation of Dirac particles}

Now we turn to investigating the quantum feature of the Kinnersley spacetime,
especially the thermal radiation of electrons, that is, we must derive the
Hawking temperature of the event horizon and the thermal radiation spectrum
of Dirac particles from the event horizon. To this end, we work out the
spinor form of Dirac equation in the Newman-Penrose (NP) \cite{NP} formalism.
We choose a complex orthogonal null-tetrad system in the Kinnersley black
hole such that its directional derivatives are
\bea
D &=& -\pr \, , ~~~~~~~~~~\Delta = \pv +G\pr \, , \nn \\
\delta &=& \frac{1}{\sqd r}\Big(-r^2W \pr +\pta
+\frac{i}{\sta}\pvi\Big) \, , \nn \\
\overline{\delta} &=& \frac{1}{\sqd r}\Big(-r^2W^* \pr
+\pta -\frac{i}{\sta}\pvi\Big) \, .
\eea
It is not difficult to determine the non-vanishing complex NP spin
coefficients in the above null-tetrad as follows \footnote{Here and
hereafter, we denote $G_{,r} = dG/dr$, etc.}
\bea
&&\rho = \frac{1}{r} \, , ~~~~\mu = \frac{G}{r} +ig\cta \, ,
~~~~\gamma = (-G_{,r} +ig\cta)/2 \, , \nn\\
&&\tau = \frac{W}{\sqd} \, , ~~~~\tilde{\pi} = -\frac{W^*}{\sqd} \, ,
~~~~\alpha = -\frac{\coa}{2\sqd r} +\frac{W^*}{\sqd} \, ,
~~~~\beta = \frac{\coa}{2\sqd r} \, , \nn\\
&&\nu = \frac{1}{\sqd r} \Big[(2rG -r^2G_{,r})W^* +r^2W_{,v}^*
+G_{,\theta} -\frac{iG_{,\varphi}}{\sta}\Big] \, .
\eea

The dynamical behavior of spin-$1/2$ particles in curved spacetime is
described by the four coupled Chandrasekhar-Dirac equations \cite{CD}
expressed in the following spinor form
\bea
&&(D +\epsilon -\rho)F_1 +(\overline{\delta} +\tilde{\pi} -\alpha)F_2
= \frac{i\mu_0}{\sqd}G_1 \, , \nn\\
&&(\Delta +\mu -\gamma)F_2 +(\delta +\beta -\tau)F_1
= \frac{i\mu_0}{\sqd}G_2 \, ,\nn\\
&&(D +\epsilon^* -\rho^*)G_2 -(\delta +\tilde{\pi}^* -\alpha^*)G_1
= \frac{i\mu_0}{\sqd}F_2 \, , \nn\\
&&(\Delta +\mu^* -\gamma^*)G_1 -(\overline{\delta} +\beta^* -\tau^*)G_2
= \frac{i\mu_0}{\sqd}F_1 \, , \label{DCP}
\eea
where $\mu_0$ is the mass of Dirac particles. Inserting the needed NP
spin coefficients and making substitutions $P_1 = \sqd rF_1$, $P_2 = F_2$,
$Q_1 = G_1$, $Q_2 = \sqd rG_2$ into Eq. (\ref{DCP}), we obtain
\bea
&&-{\cD}_0 P_1 +\big({\cL} -r^2W^*{\cD}_2\big) P_2 = i\mu_0 r Q_1 \, ,
\nn\\
&&2r^2{\cB}_1^{\dagger} P_2 +\big({\cLd} -r^2W{\cD}_0\big) P_1
= i\mu_0 r Q_2 \, , \nn\\
&&-{\cD}_0 Q_2 -\big({\cLd} -r^2W{\cD}_2\big) Q_1 = i\mu_0 r P_2 \, , \nn\\
&&2r^2{\cB}_1 Q_1 -\big({\cL} -r^2W^*{\cD}_0\big) Q_2= i\mu_0 r P_1 \, ,
\label{reDP}
\eea
in which we have defined operators
\bea
&&{\cB}n = \pv +G{\cD}_n +(G_{,r} -ig\cta)/2 \, ,
~~~~~~{\cD}_n = \pr +n/r \, , \nn\\
&&{\cL} = \pta +\frac{1}{2}\coa -\frac{i}{\sta}\pvi  \, ,
~~{\cLd} = \pta +\frac{1}{2}\coa +\frac{i}{\sta}\pvi  \, . \nn
\eea

Because the Chandrasekhar-Dirac equation (\ref{DCP}) can be satisfied by
identifying $G_1$, $G_2$ with $F_2^*$, $-F_1^*$, respectively, so one may
deal with a pair of components $(P_1, P_2)$ only. Although Eq. (\ref{reDP})
can not be decoupled, to deal with the Hawking radiation, one should be
concerned about the asymptotic behavior of Eq. (\ref{reDP}) near the horizon
only. Under the transformation (\ref{trans}), Eq. (\ref{reDP}) with respect
to ($P_1, P_2$) can be reduced to the following limiting form near the event
horizon
\bea
&&\spr P_1 +\Big(r_{H,\theta} -\frac{i}{\sta_0}r_{H,\varphi}
+r_H^2W^* \Big)\spr P_2 = 0 \, , \nn\\
&&\Big(r_{H,\theta} +\frac{i}{\sta_0}r_{H,\varphi}
+r_H^2W\Big)\spr P_1 -2r_H^2\Big(G -r_{H,v}\Big)\spr P_2 = 0 \, ,
~~~~~~~~~~\label{trDPP}
\eea
after being taken the $r \rightarrow r_H(v_0,\theta_0,\varphi_0)$,
$v \rightarrow v_0$, $\theta \rightarrow \theta_0$ and $\varphi
\rightarrow \varphi_0$ limits. If the derivatives $\spr P_1$ and
$\spr P_2$ in Eq. (\ref{trDPP}) are not equal to zero, the existence
condition of non-trivial solutions for $P_1$ and $P_2$ is that its
determinant vanishes, which gives the above-head equation (\ref{eh}).
This treatment can be thought of as another derivation of the location
of event horizon. It is interesting to note that a similar procedure
applying to another pair of components $(Q_1, Q_2)$ will bring about
the same result.

To investigate the Hawking radiation of spin-$1/2$ particles, we need
deal with the behavior of the second-order Dirac equations near the
event horizon. A direct calculation gives the second-order form of
Dirac equations for ($P_1, P_2$) components as follows
\bea \label{socd+}
&&\Big[2r^2{\cB}_0{\cD}_0 +({\cL} -r^2W^*{\cD}_{-1})({\cLd}
-r^2W{\cD}_0) -\mu_0^2 r^2 \Big]P_1 \nn\\
&&+2r^2\Big[({\cL} -r^2W^*{\cD}_1){\cB}_1^{\dagger}
-{\cB}_0({\cL} -r^2W^*{\cD}_2)\Big]P_2 = 0 \, ,
\eea
and
\bea \label{socd-}
&&\Big[2r^2{\cD}_1{\cB}_1^{\dagger} +({\cLd}
-r^2W{\cD}_1)({\cL} -r^2W^*{\cD}_2)
-\mu_0^2 r^2 \Big] P_2 \nn\\
&&+\Big[{\cD}_{-1}({\cLd} -r^2W{\cD}_0) -({\cLd}
-r^2W{\cD}_1){\cD}_0 \Big] P_1 = 0\, .
\eea

Given the GTCT in Eq. (\ref{trans}) and after some lengthy calculations,
the limiting form of Eqs. (\ref{socd+},\ref{socd-}), when $r$ approaches
$r_H(v_0, \theta_0, \varphi_0)$, $v$ goes to $v_0$, $\theta$ goes to
$\theta_0$ and $\varphi$ goes to $\varphi_0$, yields
\bea \label{wone+}
&&{\cK}P_1 +\Big[-A +r_H^2G_{,r} +2r_H^2G +2r_H^3W^*W
+2ig\cta_0 -2r_H^2f\coa_0 \nn\\
&&~~~+\coa_0\Big(-r_{H,\theta} +\frac{ir_{H,\varphi}}{\sta_0}\Big)
-\Big(r_{H,\theta\theta} +\frac{r_{H,\varphi\varphi}}{\sda_0}
\Big)\Big] \spr P_1 +2r_H^2\Big[G_{,\theta} \nn\\
&&~~~ -\frac{i}{\sta_0}G_{,\varphi} +r_H^2W_{,v}^*
+W^*(2r_HG -r_H^2G_{,r})\Big] \spr P_2 = 0 \, ,
\eea
and
\bea \label{wone-}
&&{\cK} P_2 +\Big[-A +3r_H^2G_{,r} +r_H(6G -4r_{H,v})
+6r_H^3W^*W -2r_H^2f\coa_0\nn\\
&&~~ +(4r_Hf -\coa_0)r_{H,\theta} -\Big(4gr_H
+i\coa_0\Big)\frac{r_{H,\varphi}}{\sta_0} \nn\\
&&~~ -\Big(r_{H,\theta\theta}+\frac{r_{H,\varphi\varphi}}{\sda_0}
\Big)\Big] \spr P_2 = 0 \, .
\eea
In the above, the operator stands for the term involving the
second-order derivatives
\bea
&&{\cK} = \Big[\frac{A}{2\kappa} +2r_H^2(2G -r_{H,v}) +2r_H^4W^*W
+2r_H^2(f r_{H,\theta} -g r_{H,\varphi})\Big] \spdr \nn\\
&&~~ +2r_H^2 \spdvr2 -\Big(f r_H^2 +r_{H,\theta}\Big) \spdra
+2\Big(g r_H^2 -\frac{r_{H,\varphi}}{\sda_0}\Big) \spdri \, . ~~\nn
\eea
By the event horizon equation (\ref{loca}) or (\ref{eh}), we know
that the coefficient $A$ is an infinite limit of $0/0$-type. Using
the L' H\^{o}spital rule, we arrive at its obvious result
\bea
A &=& \lim_{r \rightarrow r_H}
\frac{2r^2(G -r_{H,v}) +r^4W^*W +2r^2(f r_{H,\theta}
-gr_{H,\varphi}) +r_{H,\theta}^2
+\frac{r_{H,\varphi}^2}{\sda}}{r -r_H} \nn\\
&=& 2r_H^2G_{,r} +2r_H^3W^*W -2r_H^{-1}(r_{H,\theta}^2
+r_{H,\varphi}^2/\sda_0) \, .
\eea

Now let us select the adjustable parameter $\kappa$ in the
operator ${\cK}$ such that
$$r_H^2 \equiv \frac{A}{2\kappa} +2r_H^2(2G -r_{H,v})
+2r_H^4W^*W +2r_H^2(f r_{H,\theta} -g r_{H,\varphi}) \, ,$$
which means the surface gravity of the horizon is
\be
\kappa =\frac{r_H^2G_{,r} +r_H^3W^*W -r_H^{-1}(r_{H,\theta}^2
+r_{H,\varphi}^2/\sda_0)}{r_H^2(1-2G) -r_H^4W^*W +r_{H,\theta}^2
+r_{H,\varphi}^2/\sda_0} \, .
\ee
With such a parameter adjustment and using relations (\ref{trDPP}),
Eqs. (\ref{wone+},\ref{wone-}) can be recast into the following
standard wave equation near the horizon in an united form
\be
\Big[\spdr +2\spdvr -2C_3\spdra +2\Omega\spdri
+2(C_2 +iC_1)\spr\Big] \Psi = 0 \, , \label{wave}
\ee
where $\Omega$, $C_3$, $C_2$, $C_1$ shall be regarded as finite
real constants. The physical meaning of $\Omega$ is that it can
be interpreted as the angular velocity of the black hole due to
deformation. For completeness, they are listed as follows
$$\Omega = g -\frac{r_{H,\varphi}}{r_H^2\sda_0} \, ,
~~~~~~~~ C_3 = f +\frac{r_{H,\theta}}{r_H^2} \, ,$$
while both $C_1$ and $C_2$ are real,
\bea
&&2(C_2 +iC_1) = \frac{2G}{r_H} -G_{,r} +2ig\cta_0
-\frac{\coa_0}{r_H^2}\Big(r_{H,\theta}
-\frac{ir_{H,\varphi}}{\sta_0}\Big) \nn\\
&&~~ -2f\coa_0 +\frac{2}{r_H^3}\Big(r_{H,\theta}^2
+\frac{r_{H,\varphi}^2}{\sda_0}\Big)
-\frac{1}{r_H^2} \Big(r_{H,\theta\theta}
+\frac{r_{H,\varphi\varphi}}{\sda_0}\Big) \label{coef} \\
&&~~ +\frac{C_3 -i\Omega\sta_0}{(G -r_{H,v})r_H^2}
\Big[r_H^2W_{,v}^* +W^*(2r_HG -r_H^2G_{,r}) +G_{,\theta}
-\frac{i}{\sta_0}G_{,\varphi}\Big] ~~~~ \nn
\eea
for $\Psi = P_1$, and
\bea
&&2(C_2 +iC_1) = \frac{2G}{r_H} +G_{,r}
+2r_HW^*W -2f\coa_0 \nn\\
&&~~~~~~~~~ -\frac{\coa}{r_H^2}\Big(r_{H,\theta}
+\frac{i r_{H,\varphi}}{\sta_0}\Big)
-\frac{1}{r_H^2}\Big(r_{H,\theta\theta}
+\frac{r_{H,\varphi\varphi}}{\sda_0}\Big)
\eea
for $\Psi = P_2$. The expression in Eq. (\ref{coef}) is very
complicated, but one can notice that the last term in the
square bracket is proportional to $\nu(r_H)$ (i.e. the value
of the spin coefficient $\nu$ at the event horizon $r_H$).

Because all coefficients in Eq. (\ref{wave}) can be viewed as
constants approximately, the wave equation can be manipulated
like an ordinary differential one. Now separating variables as
$\Psi = R(r_*)\Theta(\theta_*)\Phi(\varphi_*)e^{-i\omega v_*}$
and substituting it into Eq. (\ref{wave}), one gets
\bea
&& \Theta^{\prime} = \lambda \Theta \, ,
~~~~~~\Phi^{\prime} = (\sigma +im) \Phi \, , \nn\\
&& R^{\prime\prime} = 2i(\omega -m\Omega -C_1 +iC_0) R^{\prime} \, ,
\eea
where $C_0 = C_2 -\lambda C_3 +\sigma\Omega$, $\lambda$ is a
real constant introduced in the separation variables, $\omega$
the energy of electrons, $m$ the quantum number of its azimuthal
angular momentum. The solutions are
\bea
&& \Theta = e^{\lambda \theta_*} \, ,
~~~~~~~~~~ \Phi = e^{(\sigma+im)\varphi_*}\, , \nn\\
&& R = R_1 e^{2i(\omega -m\Omega - C_1)r_* -2C_0r_*} +R_0 \, .
\eea

The ingoing wave and the outgoing wave to Eq. (\ref{wave})
are, respectively,
\bea
&&\Psi_{\rm in} = e^{-i\omega v_* +(\sigma+im)\varphi_*
+\lambda \theta_*} \, , \nn\\
&&\Psi_{\rm out} = \Psi_{\rm in}e^{2i(\omega -m\Omega
- C_1)r_* -2C_0r_*} \, , ~~~~ (r > r_H) \, .
\eea
Near the event horizon, we have $r_* \sim (2\kappa)^{-1}\ln (r - r_H)$.
Clearly, the outgoing wave $\Psi_{\rm out}(r > r_H)$ has a logarithm
singular and is not analytic at the event horizon $r = r_H$, but can
be analytically extended from the outside of the hole into the inside
of the hole through the lower complex $r$-plane (i.e., $ (r -r_H)
\rightarrow (r_H -r)e^{-i\pi}$) to
\be
\widetilde{\Psi_{\rm out}} = \Psi_{\rm in}e^{2i(\omega -m\Omega
-C_1)r_* -2C_0r_*}e^{i\pi C_0/\kappa}e^{\pi(\omega -m\Omega
-C_1)/\kappa} \, , ~~~~(r < r_H) \, .
\ee

According to the method suggested by Damour and Ruffini \cite{DR}
and developed by Sannan \cite{San}, the relative scattering
probability of the outgoing wave at the horizon is easily obtained
\be
\Big|\frac{{\Psi}_{\rm out}}{\widetilde{\Psi_{\rm out}}}\Big|^2
= e^{-2\pi(\omega -m\Omega - C_1)/\kappa} \, .
\ee
The thermal radiation spectrum of Dirac particles from the event
horizon of the hole is given by the Fermionic distribution
\be\label{sptr}
\langle {\cal N}(\omega) \rangle = \frac{\Gamma(\omega)}{e^{(\omega
-m\Omega - C_1)/T_H } + 1} \, ,
\ee
in which $\Gamma(\omega)$ is the barrier factor of certain modes,
and the Hawking temperature $T_H = \kappa/(2\pi)$ is obviously
expressed as
\be
T_H = \frac{1}{4\pi r_H} \times \frac{M r_H -r_H^3a\cta_0
-r_{H,\theta}^2 -r_{H,\varphi}^2/\sda_0}{M r_H +r_H^3a\cta_0
+2^{-1}(r_{H,\theta}^2 +r_{H,\varphi}^2/\sda_0)} \, .
\label{temp}
\ee
It follows that the temperature depends not only on the time $v$, but
also on the angles $\theta$ and $\varphi$ because it is determined by
the surface gravity $\kappa$, a function of $v, \theta$, and $\varphi$.
The temperature is consistent with that derived from the investigation
of the thermal radiation of Klein-Gordon particles in the Kinnersley
black hole \cite{LZZZ}.

\section{Spin-acceleration coupling effect}

There are two parts in the thermal radiation spectrum (\ref{sptr}), one
is the rotational energy $m\Omega$; another is $C_1$ due to the coupling
between the spin of electrons and the angular momentum of the black hole.
Comparing the thermal spectrum of spin-$1/2$ particles with that of scalar
particles \cite{LZZZ}, we find that an extra term $C_1$ appears in the
former and is absent in the latter. From its explicit expression,
\bea
C_1 &=&\frac{C_3\sta_0}{2(G -r_{H,v})}\Big[r_H^2g_{,v}
+(2r_HG -r_H^2G_{,r})g -\frac{G_{,\varphi}}{\sda_0}\Big] \nn\\
&& -\frac{\Omega\sta_0}{2(G -r_{H,v})}\Big[r_H^2f_{,v}
+(2r_HG -r_H^2G_{,r})f +G_{,\theta}\Big] \nn\\
&& +g\cta_0 +\frac{r_{H,\varphi}\cta_0}{2r_H^2\sda_0} \, ,
~~~~~~(\Psi = P_1) \label{spef+} \\
C_1 &=& -\frac{r_{H,\varphi}\cta_0}{2r_H^2\sda_0} \, ,
~~~~~~~~~~~~~~~~~~~~~~(\Psi = P_2) \label{spef-}
\eea
one can find that $C_1 = 0$ for $\Psi = P_1, P_2$ in the non-uniformly
rectilinearly accelerating black hole ($b = c = 0$ and $r_{H,\varphi}
= 0$). Further more, if we only consider the second term in Eq.
(\ref{spef+}), then we can rewrite the ``spin-dependent'' term as
\be
\omega_s \sim \frac{sr_{H,\varphi}\cta_0}{r_H^2\sda_0} \, ,
~~~~ (s = 1/2, ~ \Psi = P_1; ~~s = -1/2, ~ \Psi = P_2)
\ee
for different spin states $s = \pm 1/2$ respectively. Here we are
interested especially in this term because it is obviously related
to the spin of electrons in different helicity states. The factor
$r_{H,\varphi}$ describes the deformation of black hole during its
evolution, while the factor $\cta_0$ comes from the scalar product
between the spin vector of electrons and the ``deformation angular
momentum'' $r_{H,\varphi}/(r_H^2\sda_0)$ of the black hole. One can
notice that this ``deformation angular momentum'' is also contained
in the expression of $\Omega$. Thus this new term represents the
spin-acceleration coupling effect which can be interpreted as the
interaction between the angular variation of the black hole and
the spin of the particles.

One can observe that $\omega_s$ vanishes at the equator $\theta =\pi/2$
and diverges in $s\theta^{-2}$ near the north pole $\theta \approx 0$
and in $-s(\pi -\theta)^{-2}$ near the south pole $\theta \approx \pi$.
In general, the dependence of $C_1$ upon the mass $M$ and the acceleration
parameters ($a, b, c$) is indirect and non-trivial because they are involved
in the expression of $r_H$. To understand this point better, we consider
a particular case where we take $M$, $a$, $b$ and $c$ as constants and set
$\theta = \theta_0 = \pi/2$. In this case, we have $2G = 1 -2M/r -r^2f^2$,
$f = b\sin\varphi +c\cos\varphi -a$ and $g = 0$ as well as $r_{H,v} =
r_{H,\theta} = 0$. Thus the event horizon equation becomes
\be
r_H^2 -2Mr_H +r_{H,\varphi}^2  = 0 \, ,
\ee
and it has solutions $r_H = M(1\pm\cos\varphi)$; $M(1\pm\sin\varphi)$,
which demonstrate that the event horizon is a triangle function of
$\varphi$. Without loss of generality, let's take one solution as
$r_H = M(1 +\cos\varphi)$. Then the Hawking temperature is
$$T_H = \frac{\cos\varphi}{2\pi M(1 +\cos\varphi)(3 -\cos\varphi)} \, .$$
Further, we have $C_3 = 0$ and $\Omega = \sin\varphi/M(1 +\cos\varphi)^2$.
For $\Psi = P_2$, $C_1$ vanishes; while for $\Psi = P_1$ due to $f_{,v} =
G_{,\theta} = 0$, it becomes
\be
C_1 = \frac{\sin\varphi(\cos\varphi -2)(b\sin\varphi
+c\cos\varphi -a)}{\sin^2\varphi +M^2(1
+\cos\varphi)^4(b\sin\varphi +c\cos\varphi -a)^2} \, .
\ee
Obviously $C_1$ is linearly dependent on $a$, $b$ and $c$ if $M = 0$.
We note that $r_H = 2M$, $T_H = 1/(8\pi M)$ and $\Omega = C_1 = 0$ at
$\varphi = \varphi_0 = 0$.

In the general case where $M$, $a$, $b$ and $c$ are arbitrary functions
of the advance time $v$, to analysis $C_1$, $\Omega$ and $T_H$ is, however,
very complicated. It needs us first to determine the explicit expression
of $r_H$ in terms of these parameters, but this topic apparently exceeds
the context of this paper. We wish to discuss them in other circumstances.

\section{Conclusions}

Equations (\ref{eh}) and (\ref{temp}) give the location and the
temperature of the event horizon of the Kinnersley black hole,
which depend not only on the advanced time $v$ but also on the
angles $\theta,\varphi$. Eq.(\ref{sptr}) shows the thermal radiation
spectrum of Dirac particles in the arbitrarily accelerating Kinnersley
spacetime, in which an extra term $C_1$ appears. We find that $C_1$
vanishes when $b = c = 0$ and $r_{H,\varphi} = 0$. This means that
this term does not exist in the rectilinearly accelerating Kinnersley
black hole. We contend that this new effect probably arise from the
interaction between the spin of Dirac particles and the acceleration of
the evaporating black hole. Besides, we notice that this spin-acceleration
coupling effect appearing in the Fermionic thermal spectrum of Dirac
particles is not shown in the Bosonic spectrum of scalar particles.

In summary, we have studied the Hawking radiation of Dirac particles
in an arbitrarily accelerating Kinnersley black hole whose mass changes
with time. The location and the temperature of the event horizon of the
accelerating Kinnersley black hole are just the same as those obtained
in the discussion on thermal radiation of Klein-Gordon particles in the
same spacetime. But the thermal spectrum of Fermi-Dirac distribution of
particles with spin-$1/2$ displays an extra interaction effect between
the spin of Dirac particles and the angular acceleration of black holes.
The character of this term is its obvious dependence on the different
spin states. This spin-acceleration coupling effect is absent from the
thermal radiation spectrum of Klein-Gordon particles. In addition, our
discussion here is easily generalized to the most general case of a
non-uniformly accelerating Kinnersley black hole with electric charge
$Q(v)$, magnetic charge $P(v)$ and cosmological constant $\Lambda$
($2G = 1 -2M(v)/r +(Q^2 +P^2)/r^2 -4a\cta(Q^2 +P^2)/r -2a r\cta
-r^2(f^2 +g^2\sda) -\Lambda r^4/3$).

\section*{Acknowledgment}

This work was supported in part by the NSFC in China. We especially thank
our referee for his several advice on improving this manuscript and for
clarifying the incorrect statement on the observational possibility of the
spin-acceleration coupling effect in astrophysics.

\end{document}